\documentclass[12pt]{article}

\usepackage{amssymb}\usepackage{bm}\usepackage{amsfonts}
\newcommand{\beq}{\begin{equation}}\newcommand{\eeq}{\end{equation}}\newcommand{\beqa}{\begin{eqnarray}}
\newcommand{\eeqa}{\end{eqnarray}}\newcommand{\w}{\wedge}\newcommand{\ts}{\textstyle}
\newcommand{\dl}{\bm{\delta}}\newcommand{\nn}{\nonumber}\newcommand{\vep}{\varepsilon}
\newcommand{\bOmega}{\bm{\Omega}}
\newcommand{\PS}{P_{\star}}
\begin{document}
{\renewcommand{\thefootnote}{\fnsymbol{footnote}}
\hfill IGC--08/5--2\\
\medskip
\begin{center}
{\LARGE  A New Perspective on Covariant Canonical Gravity}\\
\vspace{1.5em}
Andrew Randono\footnote{e-mail address: {\tt arandono@gravity.psu.edu}}
\\
\vspace{0.5em}
Institute for Gravitation and the Cosmos,\\
The Pennsylvania State
University,\\
104 Davey Lab, University Park, PA 16802, USA\\
\vspace{1.5em}
\end{center}
}

\setcounter{footnote}{0}

\abstract{We present a new approach to the covariant canonical formulation of Einstein-Cartan gravity that preserves the full Lorentz group as the local gauge group. The method exploits lessons learned from gravity in 2+1 dimensions regarding the relation between gravity and a general gauge theory. The dynamical variables are simply the frame field and the spin-connection pulled-back to the hypersurface, thereby eliminating the need for simplicity constraints on the momenta. A consequence of this is a degenerate (pre)symplectic form, which appears to be a necessary feature of the Einstein-Cartan formulation. A new feature unique to this approach arises when the constraint algebra is computed: the algebra is a deformation of the de Sitter, anti-de Sitter, or Poincar\'{e} algebra (depending on the value of the cosmological constant) with the deformation parameter being the conformal Weyl tensor.}

\section{Introduction}
The similarities between general relativity, a theory about metric dynamics, and general gauge theory, a theory about dynamical connections, is nowhere more apparent than in the Einstein-Cartan formulation of gravity and extensions therein. Indeed, in 2+1 dimensions in the presence of a cosmological constant, the theory can be reformulated precisely in the form of a gauge theory, namely Chern-Simons theory of an $\mathrm{(A)dS}_3$ connection \cite{Carlip:Book}. However, in four dimensions, which will be the concern of this paper, there are essential difference that distinguish gravity from ordinary gauge theories. Most notably, whereas the dynamical ingredients in a gauge theory are typically the connection coefficients alone, often on a fixed background, in the Einstein-Cartan theory the dynamical ingredients are a Lorentz connection together with a dynamical frame field. Although the Macdowell-Mansouri formulation, just as in 2+1 dimensions, repackages the connection and the frame field into a single de Sitter, anti-de Sitter, or Poincar\'{e} connection (depending on the value of the cosmological constant), the full gauge invariance is necessarily broken at the level of the action \cite{MMoriginal,Wise:MMgravity}. When the canonical analysis of Einstein-Cartan theory is carried out, the initial phase of the construction is promising---the theory appears to have the same structure as Yang-Mills theory with the Lorentz connection playing the role of the gauge field, and a particular combination of the tetrad playing the role of the conjugate momentum. However, a full treatment of the constraint algebra unveils the existence of second class constraints, the most natural (though not the only) implementation of which requires the breaking of the Lorentz group down to its rotation subgroup, and once this is carried out the dynamical fields no longer consist of a connection and its conjugate momentum \cite{Ashtekar:book}. To circumvent this, as is done in Loop Quantum Gravity, one can add a parity violating term to the action which does not affect the classical equations of motion, thereby regaining a connection as a dynamical variable \cite{Holst, Barbero}. However this connection is not a connection of the full Lorentz group but is strictly a three-dimensional connection of the rotation subgroup \cite{Rovelli:book}. This situation has been slightly improved recently in \cite{RovelliFatFranc} where it is shown that a connection can be defined globally on the spacetime such that the pull-back of this connection is the Ashtekar-Barbero-Immirzi connection. However, the construction of the global connection requires a projection of the Lorentz group as an intermediate step, and as a consequence it can no longer be interpreted geometrically as a spin--connection. Taking a more covariant approach, one can complexify the gauge group and consider the dynamics of only the left or right-handed components of the spin connection whose pull-backs to the hypersurface are the Ashtekar variables \cite{Ashtekar:book}. Then the canonical formalism does bear a striking similarity to a standard gauge theory, however this is obtained at the expense of complexification---decomplexification requires the implementation of complicated reality constraints not familiar to standard gauge theories. Another approach, the approach underlying covariant Loop Quantum Gravity \cite{Alexandrov:Covariant, Livine:Covariant}, retains full Lorentz covariance without complexification by implementing primary ``simplicity" constraints on some of the dynamical variables, but there again the constraint algebra reveals the emergence of second class constraints which must be dealt with. Upon implementation, these constraints effectively modify the dynamical coefficients such that the components themselves no longer commute under the Poisson bracket.

Any way one tries to manipulate it, despite the undeniable similarity with gauge theories, general relativity resists interpretation as an ordinary gauge theory based on the local Lorentz group.

In this work, we provide a new perspective on the (real) covariant formulation of canonical gravity. In particular we develop an approach that exploits the similarities between Einstein-Cartan theory and a gauge theory based on the Lorentz group, while highlighting the essential differences between the approaches. Our approach will take many of the lessons learned from the canonical formulation of gravity in 2+1 dimensions, and attempt to carry these over to four dimensions. In particular, the dynamical variables will be simply the pull-back of the spin connection and frame field to the hypersurface, thereby avoiding the need for simplicity constraints on the dynamical variables. The expense to be paid is a degeneracy in the now pre-symplectic structure, which is a characteristic (and essential) feature of the four dimensional theory. As in 2+1 dimensions, the Hamiltonian constraint is a vectorial constraint (in contrast to the existing covariant approaches) whose vector generators are closely related to the generators of translations of the (A)dS or Poincar\'{e} group. In 2+1 dimensions, since the theory is topological the constraint algebra is precisely the (A)dS/Poincar\'{e} algebra. Despite the degeneracy in the symplectic structure of the four dimensional theory, we find we are able to compute the constraint algebra. A new feature, unique to this approach, emerges: as a consequence of the local degrees of freedom of general relativity, the algebra is a deformation of the (A)dS/Poincar\'{e} algebra with the conformal Weyl tensor playing the role of the deformation parameter. This very clearly illustrates precisely how general relativity is similar to a gauge theory (as in 2+1 dimensions) and how it differs.

Though we will not explore these issues in this paper, as a practical side benefit, since the the dynamical variables are simply the pull-back of the spin connection and frame field, this approach may be more suitable for coupling spinors to the gravitational field than the standard approach where the simplicity constraints on the conjugate momenta severely complicate the theory \cite{Bojowald:Fermions}. In addition, we hope this work will provide insight into the origin of the second class constraints in covariant Loop Quantum Gravity, and the reason for the con-commutativity of the connection coefficients.

Though we hope this paper is self-contained, it is a follow-up to \cite{Randono:CanonicalLagrangian} where we develop some new machinery for dealing the Hamiltonian construction of general relativity in a covariant manner. We will briefly overview the major concepts, but we refer the reader to \cite{Randono:CanonicalLagrangian} for more details.

\section{Time evolution and the (A)dS/Poincar\'{e} algebra}
Since the (A)dS/Poincar\'{e} algebra, will play an important role in what follows begin by briefly reviewing the relevant algebras and heuristically explaining their relation to the time evolution of gravity in vacuum.

We will adopt most of the notation in \cite{Randono:CanonicalLagrangian}. It will be convenient to adopt a Clifford algebra notation to facilitate the relation between the constraint algebra and the (A)dS/Poincar\'{e} algebra. As usual, the Clifford algebra is defined by the fundamental equation $\gamma^I \gamma^J+\gamma^J\gamma^I=2\eta^{IJ}$,
where $\eta^{IJ}=diag(-1,1,1,1)$ is the canonical bi-linear form of Lorentzian signature. The Clifford algebra is the algebra formed by linear combinations of products of the $\gamma^I$ matrices.
 The (A)dS/Poincar\'{e} group is a ten-dimensional group (in four spacetime dimensions) isomorphic to the set of isometries of de
Sitter, anti-de Sitter, or Minkowski spacetime. The algebra consists of the Lorentz generators
$\lambda=\frac{1}{4}\gamma_{I}\gamma_{J}\,\lambda^{[IJ]}$, and the pseudo-translation generators
$\eta=\frac{i}{2\epsilon r_{0}}\,\gamma_{K}\eta^{K}$, where $\epsilon=1$ for the de Sitter
algebra, and $\epsilon=i$ for the anti-de Sitter alegbra. For convenience we have
included the cosmological constant in the algebra itself via $r_{0}\equiv \sqrt{\frac{3}{|\Lambda|}}$. The de Sitter
algebra then follows from the Clifford algebra:
\beqa
\left[\lambda_{(1)}, \lambda_{(2)}\right]&=&\ts{\frac{1}{4}}\gamma_{[I}\gamma_{J]}\,\lambda^{I}_{(1)\,K}\lambda^{KJ}_{(2)} \\
\left[\lambda, \eta\right]&=&\ts{\frac{i}{2\epsilon r_{0}}}\gamma_{I} \,{\lambda^{I}}_{K}\eta^{K}\\
\left[\eta_{(1)},\eta_{(2)}\right]&=&
-\ts{\frac{\Lambda}{3}}\,\ts{\frac{1}{4}}\gamma_{I}\gamma_{J}\,\eta^{[I}_{(1)}\eta^{J]}_{(2)}\,.
\eeqa

The fundamental ingredients in the Einstein-Cartan formulation of general relativity are an orthonormal frame field\footnote{Strictly speaking, the set is a co-frame field since it is a set of one-forms, but we will be loose and henceforth refer to them as frame fields. The true vector frame will be denoted by a different symbol, $\bar{\vartheta}_I$, so there will be no confusion.}\footnote{By orthonormal, we mean here that the frame is orthonormal with respect to the inverse of the induced metric $g\equiv \eta_{IJ}\,\vep^I \otimes \vep^J$ so that $g^{-1}(\vep^I, \vep^J)=\eta^{IJ}$. This is somewhat of a tautology given that the frame is assumed to be invertible and the induced metric is {\it defined} by the frame itself. In this respect, the qualifier ``orthonormal" simply distinguishes this type of frame from, say, the commonly used {\it null}--tetrad, which do not share this orthogonality property with respect to their induced metric}, $\vep\equiv
\frac{1}{2}\gamma_{I}\,\vep^{I}_{\mu}dx^{\mu}$, here valued in the vector elements of the Clifford algebra, and a $Spin(3,1)$ connection that, in a local trivialization of the bundle, can be represented by the $\mathfrak{spin}(3,1)$ valued one-form
$\varpi\equiv \frac{1}{4}\gamma_{I}\gamma_{J}\,{\varpi^{[IJ]}}_{\mu}dx^{\mu}$, where we have chosen a Clifford algebra representation of $\mathfrak{spin}(3,1)$. The global situation can be captured in the language of principle g-bundles and their associated vector bundles (see e.g. \cite{Kobayashi, FatFranc, Frankel}), and such an analysis is likely to yield interesting topological properties, however, this is outside of the scope of this paper. Thus, we will make no assumptions about the global properties of the frame field and the global topology of the fiber-bundle, and focus on objects and expressions defined in a local trivialization.

The general solution to the Einstein-Cartan field equations in vacuum takes the form
\beqa
R_{\varpi}&=&\ts{\frac{\Lambda}{3}}\,\vep\w\vep+\mathcal{C} \nn\\
T&\equiv& D_{\varpi}\vep =0 \label{ECSolution}
\eeqa
where $R_{\varpi}=d\varpi+\varpi \w\varpi$, $D_{\varpi}$ is the covariant exterior derivative with respect to $\varpi$, $\Lambda$ is the cosmological constant, and $\mathcal{C}=\frac{1}{4}\gamma_{I}\gamma_{J}\,{\mathcal{C}^{[IJ]}}_{\mu\nu}\, \ts{\frac{1}{2}}dx^{\mu}\w dx^{\nu}$ is the conformal Weyl tensor. Choosing a timelike vector field $\bar{t}$, we contract the above equations to obtain the time evolution of the fundamental variables:
\beqa
\delta_t \varpi &=& \mathcal{L}_{\bar{t}}\,\varpi= -D_{\varpi}\lambda+\ts{\frac{\Lambda}{3}}[t,\vep]+\mathcal{C}(\bar{t},\ )\nn\\
\delta_t\vep&=&\mathcal{L}_{\bar{t}}\,\vep = [\lambda,\vep]+D_{\varpi}t
\eeqa
where we have defined the parameters $\lambda\equiv -\varpi(\bar{t})$ and $t\equiv \vep(\bar{t})$. From the above, it is evident that a large portion of the evolution is simply an (A)dS/Poincar\'{e} gauge transformation. By this we mean the following: the variables $\varpi$ and $\vep$ can be combined \`{a} la Macdowell and Mansouri \cite{MMoriginal} into a single (A)dS/Poincar\'{e} connection given by $\xi=\varpi+\frac{i}{\epsilon r_0}\,\vep$. Under an infinitesimal gauge transformation generated by the arbitrary parameter $\alpha=\lambda-\frac{i}{\epsilon r_0}t$, the connection transforms according to $\xi\rightarrow \xi -D_{\xi}\alpha$. This implies the transformations $\varpi\rightarrow \varpi -D_{\varpi}\lambda +\ts{\frac{\Lambda}{3}}[t,\vep]$, and $\vep\rightarrow \vep + [\lambda,\vep]+D_{\varpi}t$. Thus, we see that a ``large portion" of the time evolution $\varpi\rightarrow \varpi+\delta_t\varpi$ and $\vep\rightarrow \vep +\delta_t\vep$ can be identified with an (A)dS/Poincar\'{e} gauge transformation. Heuristically, we can think of the ``gauge" part of the evolution as being the strictly (A)dS/Poincar\'{e} transformation of the evolution and the ``non-gauge" part of the evolution as being simply a translation of $\varpi$ by $\mathcal{C}(\bar{t},\ )$. 

Now consider the commutator of two succesive time evolutions along different vector fields $\bar{t}_1$ and $\bar{t}_2$ given by $\delta_{t_2}\delta_{t_1}\varpi-\delta_{t_1}\delta_{t_2}\varpi=\mathcal{L}_{[\bar{t}_2,\bar{t}_1]}\varpi$ and similarly for $\vep$. Generically, one might expect that the ``gauge" part of the transformation would simply be an (A)dS/Poincar\'{e} transformation generated by $\alpha_{12}=\lambda_{12}+t_{12}=[\lambda_1+t_1, \lambda_2+t_2]$, and the ``non-gauge" part would be a translation of $\varpi$ by $\mathcal{C}([\bar{t}_2, \bar{t}_1],\ )$. The ``gauge" part of the evolution would then be a reflection of the (A)dS/Poincar\'{e} algebra. Let's see what happens. On the solution space defined by (\ref{ECSolution}), we find the transformations take the form:
\beqa
\mathcal{L}_{[\bar{t}_2,\bar{t}_1]}\,\varpi&=& -D_{\varpi}\widetilde{\lambda}+\ts{\frac{\Lambda}{3}}[\widetilde{t},\vep]+\mathcal{C}([\bar{t}_2,\bar{t}_1],\ )\nn\\
\mathcal{L}_{[\bar{t}_2, \bar{t}_1]}\,\vep &=& [\widetilde{\lambda},\vep]+D_{\varpi}\widetilde{t}
\eeqa
where
\beqa
\widetilde{t}&\equiv&\vep([\bar{t}_{2},\bar{t}_{1}])=[\lambda_{1},t_{2}]-[\lambda_{2},
t_{1}]-t^{[12]} \nn\\
\widetilde{\lambda}&\equiv&-\varpi([\bar{t}_{2},\bar{t}_{1}])=[\lambda_{1},\lambda_{2}]-\ts{\frac{\Lambda}{3}}[t_1,t_2]-\mathcal{C}(\bar{t}_{1},\bar{t}_{2})
-\lambda^{[12]}\,,
\eeqa
where $t^{[12]}\equiv \mathcal{L}_{\bar{t}_1}t_2-\mathcal{L}_{\bar{t}_2}t_1$ and $\lambda^{[12]}\equiv \mathcal{L}_{\bar{t}_1}\lambda_2-\mathcal{L}_{\bar{t}_2}\lambda_1$. Typically when computing the constraint algebra it is assumed that the Lagrange multipliers (here $t$ and $\lambda$) are non-dynamical and therefore effectively time invariant, so for the purposes of comparison, we will set $t^{[12]}$ and $\lambda^{[12]}$ equal to zero. The ``non-gauge" part of the evolution remains the same: a translation of $\omega$ by $\mathcal{C}([\bar{t}_2, \bar{t}_1],\ )$ as expected. Comparing with the (A)dS/Poincar\'{e} algebra, most of the ``gauge" part of the evolution is precisely what one would expect from the commutator of gauge transformations in the respective algebra---however, there is an anomalous term in the algebra given by the addition of the term $\mathcal{C}(\bar{t}_{1},\bar{t}_{2})$ to the $Spin(3,1)$ generators. In this sense, the ``gauge" part of the evolution is deformed from the pure (A)dS/Poincar\'{e} algebra by the presence of the conformal Weyl tensor. The major result of this paper is that when the Hamiltonian analysis is approached in a way that preserves the covariance of the Einstein-Cartan formulation of general relativity, the (A)dS/Poincar\'{e} algebraic structure and the deformation therein by the conformal Weyl tensor is reflected in the constraint algebra itself.

\section{Review of symplectic dynamics without gauge fixing}
In this section we review the construction of the Hamiltonian formulation of Einstein-Cartan gravity that retains the full Lorentz group as the local gauge group. Our construction closely follows \cite{Randono:CanonicalLagrangian}. Per usual, we assume the spacetime, $M^{*}$ is homotopic to
$\mathbb{R}\times \Sigma$ where $\Sigma$ is a typical spacelike Cauchy slice of the spacetime. This implies a restriction on the allowed field conifgurations. In particular, it restricts one to spacetimes that admit a Cauchy slicing, where the induced metric on the initial data surface is spacelike. Although the latter assumption is likely not necessary in a covariant framework, we will keep the assumption in order to make contact with the standard Hamiltonian formalism and to avoid subtleties in analysis that are outside the scope of this paper. In $M^{*}$, we embed a
spacelike hypersurface, via the embedding map $\sigma:\Sigma\rightarrow M^{*}$. The dynamical arena we are concerned
with is the submanifold $M$ that consists of all point in the past of $\Sigma$ and contains $\Sigma$ on its boundary
so that $\partial M=\Sigma$. The full spacetime is reconstructed by advancing $\Sigma$ in time. It will be important to distinguish the Lagrangian variables from their pull-backs
to the spacelike Cauchy slice so we will denote these objects by two different symbols. The pull-back fields of the dynamical variables are denoted by $\phi^{*}\vep\equiv e$ and $\phi^{*} {\varpi}\equiv
\omega$. We will assume throughout that the
tetrad are invertible. Thus, for any $\vep^{I}$ there exists a canonical dual, $\bar{\vartheta}_{J}$, such that
$\vep^{I}(\bar{\vartheta}_{J})=\delta^{I}_{J}$. This restriction implies restrictions on the three dimensional
connection which we summarize in appendix \ref{Inverses}. 

We begin with a slight generalization of the Einstein-Cartan action to include a cosmological constant and a parity
violating term known as the Holst modification \cite{Holst}. This modification classically has no effect in vacuum, but it is
necessary in Loop Quantum Gravity. The Holst modified action is
\beq
S=\int_{M}P_{\star}\vep\,\vep\,R_{\varpi}-\ts{\frac{\Lambda}{6}}\star\vep\,\vep\,\vep\,\vep \,,
\eeq
where, to avoid proliferation of symbols, we have dropped explicit wedge products and the trace over the Clifford algebra. The
element, $\star=-i
\gamma_{5}=\frac{1}{4!}\epsilon_{IJKL}\gamma^{I}\,\gamma^{J}\,\gamma^{K}\,\gamma^{L}$,
is the duality operator in the internal $Spin(3,1)$ representation space, and 
\beq
P_{\star}\equiv \star+\frac{1}{\beta}
\eeq
 where $\beta$ is the Immirzi parameter. This operator is invertible unless $\beta$ takes on the special values $\pm i$ where it becomes proportional to a chiral projection operator. For simplicity, in this paper we will always assume the operator is invertible. Those readers who are unfamiliar with the Immirzi parameter can always safely take the limit as $\beta\rightarrow \infty$, thereby replacing $\PS$ with the internal duality operator, $\star$, in every expression.

Now, consider an arbitrary variation of the action. We will treat the variation $\dl$ 
as the exterior derivative on the infinite dimensional Lagrangian phase space. Vectors and forms in the infinite dimensional Lagrangian or 
Hamiltonian phase space will be written in \textbf{bold} font. Throughout this paper we will work with a function space at a rather formal level. In a more rigorous treatment, one would precisely define the function space such that the phase space becomes a true (infinite-dimensional) differentiable manifold. Then one could rigorously define the tangent and cotangent bundles on which the vectors and forms reside. Alternatively, one could use jet bundles \cite{FatFranc,Saunders:Jet}, which resort to equivalence classes of truncations of functions in order to construct finite dimensional fiber-bundles. It is fully expected that one could perform the analysis with all due rigor, however in the interest of brevity and to maintain the specific focus of this paper, we opt for a more physics friendly language.

An arbitrary variation of the action splits into a boundary
term, $\bm{J}$, and a bulk term, $\bm{\theta}$, which we will refer to as the symplectic and Lagrangian one forms,
respectively:
\beq
\dl S=\bm{J}+\bm{\theta}
\eeq
where
\beqa
\bm{J}&=&\frac{1}{k}\int_{\partial M}P_{\star} e\, e\, \dl \omega \label{Symplectic1form}\\
\bm{\theta}&=&\frac{1}{k} \int_{M}-D(P_{\star}\vep\,\vep)\, \dl\omega +\left(P_{\star}R\, \vep -\vep\,P_{\star}R-
{\ts\frac{2\Lambda}{3}}\star \vep\,\vep\,\vep \right)\dl \vep \label{Lagrangian1form}
\eeqa
The vanishing of $\bm{\theta}$ yields the bulk equations of motion:
\beqa
P_{\star}R\,\vep-\vep\,P_{\star}R-{\ts\frac{2\Lambda}{3}}\star \vep\,\vep\,\vep&=&0 \label{EOM1}\\
D(P_{\star}\vep\,\vep)&=& 0 \,.\label{EOM2}
\eeqa
The (pre)symplectic two-form $\bm{\Omega}$ is obtained by the exterior derivative of the Lagrangian or Hamiltonian one
form:
\beqa
\bm{\Omega}&=&-\dl \bm{J} = \dl \bm{\theta}\\
&=&\frac{1}{k}\int_{\Sigma}P_{\star}\, \dl\omega \w \dl (e\, e)  \, .
\eeqa

To obtain general relativity from Hamilton's equations, we first need the total Hamiltonian constraint, which can be constructed from $\bm{\theta}$. Defining the time evolution
vector field
\beq
\bm{\bar{t}}=\int_{M}\mathcal{L}_{\bar{t}}\,\varpi\, \frac{\dl}{\dl \varpi}
+\mathcal{L}_{\bar{t}}\,\vep \, \frac{\dl}{\dl \vep}
\eeq
the total Hamiltonian constraint is
\beq
C_{tot}(\lambda, t)\equiv -\bm{\theta}(\bm{\bar{t}})=C_{G}(\lambda)+C_{H}(t)\,.
\eeq
Here $C_{G}(\lambda)$ is the Gauss constraint,
\beq
C_{G}(\lambda)=\frac{1}{k}\int_{\Sigma}-D\lambda \, P_{\star}\, e\, e
\eeq
where $\lambda\equiv -\varpi(\bar{t})$, and $C_{H}(t)$ is the Hamiltonian constraint,
\beq
C_{H}(t)= \frac{1}{k}\int_{\Sigma} -[t, e]\,
\left(P_{\star} R -\ts{\frac{\Lambda}{3}}\star e\, e\right)\,,
\eeq
where $t\equiv \vep(\bar{t})$. A peculiarity of this formalism is that the Hamiltonian constraint is not a scalar constraint, rather, it is {\it vectorial}. This will be explained when we compute the constraint algebra explicitly. We will see that the vector generators of the
constraint are directly related to the generators of (A)dS/Poincar\'{e} translations.

Hamilton's equations of motion can be written succinctly by
\beq
\bm{\mathcal{L}_{\bar{t}}}\,\bm{\theta}=0\,.
\eeq
This equation splits into two parts equivalent to
\beqa
C_G(\lambda)&\approx& C_H(t) \approx 0 \label{HAM1}\\
\bm{\Omega(\bar{t},\ )}&=&\dl C_{tot}(\lambda,t)\Big|_{\lambda,t} \label{HAM2}
\eeqa
where the vertical line in the second equation indicates that the functional derivative is taken holding $\lambda$ and $t$ fixed.
The vanishing of the constraints (\ref{HAM1}) can be shown to be pointwise identical to the pull-back of the Einstein-Cartan equations to $\Sigma$,
\beqa
\sigma^* \left(P_{\star}R_{\varpi}\,\vep-\vep\,P_{\star}R_{\varpi}-{\ts\frac{2\Lambda}{3}}\star \vep\,\vep\,\vep\right)&=&0 \nn\\
\sigma^*\left(D_{\varpi}(P_{\star}\vep\,\vep)\right)&=& 0 \label{HEOMa}\,,
\eeqa
and the Hamilton's evolution equation (\ref{HAM2}) is equivalent to the pull-back of the time components of the Einstein-Cartan equations of motion (here $i_{\bar{t}}\alpha =\alpha(\bar{t},\ ,...)$ is the insertion of $\bar{t}$ into a differential form $\alpha$),
\beqa
\sigma^* \left(i_{\bar{t}}\left(P_{\star}R_{\varpi}\,\vep-\vep\,P_{\star}R_{\varpi}-{\ts\frac{2\Lambda}{3}}\star \vep\,\vep\,\vep\right)\right)&=&0 \nn\\
\sigma^*\left(i_{\bar{t}}\left(D_{\varpi}(P_{\star}\vep\,\vep)\right)\right)&=& 0 \label{HEOMb}\,.
\eeqa
Together, these equations form the full set of the Einstein-Cartan field equations restricted to the spatial hypersurface. 

\subsection{Working with the degenerate symplectic form}
The symplectic form presented in the previous section is technically only pre-symplectic. That is, the two-form,
although anti-symmetric and closed, is degenerate. This means that the form is not invertible, and their exist
non-trivial vectors $\bm{\bar{Z}}$ such that
\beq
\bm{\Omega}(\bm{\bar{Z}},\ )=0\,.
\eeq
This degeneracy is not simply a technical nuisance, rather, it is an essential part of the Einstein-Cartan formulation
of gravity. In this section, we summarize some important facts to keep in mind when working with the degenerate
(pre)symplectic form (for more on pre-symplectic forms, see \cite{AABombelli:CovariantSymplectic,AshtekarMagnon,Gotay:MomentumMaps1,Gotay:MomentumMaps2}). In appendix \ref{AppendixB} we characterize the degeneracy further.

The two most important problems associated with the degeneracy are the following:
\begin{itemize}
\item{\textbf{Canonical vector fields do not always exist:} Given a functional $f$, it is not always true that there
exists a vector field $\bm{\bar{X}_{f}}$ such that $\bm{\Omega}(\bm{\bar{X}_{f}},\ )=\dl f$. Supposing there is such a
vector field, then we must have $\mathcal{L}_{\bm{\bar{Z}}}f=0$ for all vector fields $\bm{\bar{Z}}$ in the kernel of
$\bm{\Omega}$. Thus, the functionals with associated canonical vector fields defined on some appropriate subspace of
the phase space must be constant along the integral curves of vectors in the kernel of $\bm{\Omega}$,
which is obviously not true for every functional.}
\item{\textbf{Canonical vector fields are not unique:} Even when a functional can be paired with a canonical vector
field, the pairing is not unique. Assuming $\bm{\bar{X}}_{f}$ is the canonical vector field associated with $f$, the
vector field $\bm{\bar{X}'}_{f}=\bm{\bar{X}}_{f}+\bm{\bar{Z}}$ is also a good canonical vector field for any
$\bm{\bar{Z}}$ in the kernel of $\bm{\Omega}$ since $\bm{\Omega}(\bm{\bar{X}}_{f},\ )=\bm{\Omega}(\bm{\bar{X}'}_{f},\
)=\dl f$.} 
\end{itemize}
With regards to the first problem, as was shown in \cite{Randono:CanonicalLagrangian}, the Hamiltonian
constraint submanifold of interest is precisely the submanifold where the constraints vanish {\it and} there exists a
canonical
vector field associated with $C_{tot}(\lambda,t)$. More specifically, the constrained Hamiltonian phase space,
$\bar{\Gamma}_{H}$
is an intersection, $\bar{\Gamma}_{H}=\Gamma^{(1)}_{H}\cap \Gamma^{(2)}_{H}$ where $\Gamma^{(1)}_{H}$ is the
submanifold where the total Hamiltonian constraint vanishes and $\Gamma^{(2)}_{H}$ is the submanifold where there are canonical vector fields associated with the total Hamiltonian. As
usual, on this submanifold, the associated vector field
is identified with the time evolution vector field to yield Hamilton's equations of motion, which we have shown yield precisely the Einstein-Cartan equations on the spatial hypersurface given by (\ref{HEOMa}) and (\ref{HEOMb}). For our purposes, this will
be sufficient to define the constraint algebra.

With regards to the second problem, despite the ambiguity in
the definitions of the associated canonical vector fields, as long as the vector fields exist in a neighborhood of a
point in the phase space, the Poisson bracket between two functionals is unique at that point. To see this, suppose we
have an open subspace $U\subset \Gamma_{H}$ and there exist vector fields such that
$\bm{\Omega}(\bm{\bar{X}}_{f},\ )=\dl f$ and $\bm{\Omega}(\bm{\bar{X}}_{g},\ )=\dl g $ everywhere in $U$. Given
another vector field $\bm{\Omega}(\bm{\bar{X}'}_{f},\ )=\dl f$, the difference $\bm{\Delta
\bar{X}_{f}}=\bm{\bar{X}'}_{f}-\bm{\bar{X}}_{f}$ is in the kernel of $\bm{\Omega}$. Thus, defining the Poisson
bracket in the usual way, we have
\beq
\{f,g\}\equiv \bm{\Omega}(\bm{\bar{X}}_{f}, \bm{\bar{X}}_{g})=\bm{\Omega}(\bm{\bar{X}'}_{f}, \bm{\bar{X}'}_{g})\,.
\eeq
Thus, the Poisson bracket is unique. Furthermore, it is still true that $\{f,g\}=-\mathcal{L}_{\bm{\bar{X}}_{f}}g
=\mathcal{L}_{\bm{\bar{X}}_{g}}f$, and $\bm{\Omega}([\bm{\bar{X}}_{g},\bm{\bar{X}}_{f}],\ )=\dl\{f,g\}$. These
properties will be sufficient to allow us to compute the constraint algebra.

It will be useful to characterize the degeneracy more explicitly. Assuming there is a canonical vector field
associated with $f$, we write the vector field in component notation 
$\bm{\bar{X}}_{f}=\int_{\Sigma}X^{(\omega)}_{f}\, \frac{\dl}{\dl \omega}+X^{(e)}_{f}\, \frac{\dl}{\dl e}$, 
the definition $\bm{\Omega}(\bm{\bar{X}}_{f},\ )=\dl f$ reduces to a set of pointwise equations:
\beqa
(e\, P_{\star} X^{(\omega)}_{f}
+P_{\star} X^{(\omega)}_{f}\, e) &=& k\,\frac{\delta f}{\delta e} \label{CanVect(w)}\\
P_{\star}(X^{(e)}_{f}\, e +e\, X^{(e)}_{f}) &=& -k\,\frac{\delta f}{\delta \omega} \label{CanVect(e)}\,.
\eeqa
Thus, we see explicitly that the components $X^{(\omega)}_{f}$ and $X^{(e)}_{f}$ are only determined up to the particular
combination of variables given above. Since we have assumed the tetrad is invertible,
the second equation can be solved uniquely for the components $X^{(e)}_{f}$ as elaborated in appendices \ref{Inverses}
and \ref{AppendixB}.
However, the first equation cannot
be inverted to give $X^{(\omega)}_{f}$. An explicit calculation reveals that this equation reduces the local degrees of
freedom of $X^{(\omega)}_{f}$ from eighteen down to six. Thus, the local dimension of the kernel of $\bm{\Omega}$ is
six. This will be important when counting the local degrees of freedom of the theory.

\section{The Constraint Algebra}
The fact that the components of the canonical vector fields are not determined uniquely presents problems in 
computing the commutator of two arbitrary generating functionals. The commutator between two functionals is well defined only on submanifolds where canonical vector fields associated with the functionals exist. We are primarily interested in the the constraint algebra so we will restrict our attention to expressions involving the Gauss and Hamiltonian constraint. It can easily be shown that the Gauss constraint is everywhere constant along the integral curves of vectors in the kernel of $\bOmega$. That is, for any $\bm{\bar{Z}}$ such that $\bOmega(\bm{\bar{Z}},\ )=0$, we have
\beqa
\bm{\mathcal{L}_{\bar{Z}}}\,C_G(\lambda)=0\,.
\eeqa
Thus, equations (\ref{CanVect(e)}) and (\ref{CanVect(w)}) with $f$ the Gauss constraint can be solved for at every point of the unconstrained Hamiltonian phase space and the resulting canonical vector field is given by (up to addition of a vector in the kernel of $\bOmega$):
\beq
\bm{\bar{\lambda}}=\int_{\Sigma}-D\lambda \,\frac{\dl}{\dl \omega}+[\lambda, e]\,\frac{\dl}{\dl e}\,.
\eeq
Because of this the commutator of $C_G(\lambda)$ with any other functional $f$ can be computed on any submanifold where $\bm{\bar{X}_f}$ is defined.
As mentioned previously, the existence of canonical vector fields associated with the total Hamiltonian is an integral part of the Einstein equations, and this requirement defines a non-zero\footnote{It can easily be shown that, for example, any submanifold of $\Gamma_H$ where the pull-back of the torsion to the hypersurface vanishes identically is contained in $\Gamma^{(2)}_H$, which demonstrates that $\Gamma^{(2)}_H$ is not empty.} submanifold $\Gamma^{(2)}_H$ in addition to the usual constraint submanifold. Thus, in order to have a well-defined constraint algebra, we must assume the existence of canonical vector fields, and we will do so throughout this section. On this submanifold, the constraint algebra is well-defined and can be computed. 

The Gauss--Gauss and Gauss--Hamiltonian commutators are
relatively straightforward to compute. The commutator of two Hamiltonian constraints
will require a bit more work, but it also can be computed using this method. Aside from this commutator, the constraint algebra can be computed to give:
\beqa
\{C_{G}(\lambda_{1}), C_{G}(\lambda_{2})\} &=& C_{G}([\lambda_{1}, \lambda_{2}])\nn \\
\{C_{G}(\lambda), C_{H}(t)\} &=& C_{H}([\lambda, t]) \nn\\ 
\{C_{H}(t_{1}),C_{H}(t_{2})\}&=& ?? \label{Commutator1}
\eeqa
We have already obtained a result different from the standard constraint algebra of general relativity. In particular, in the standard approach, the commutator between the Gauss and Hamiltonian constraint vanishes identically. The fact that the commutator above does not vanish is a consequence of the vectorial nature of the Hamiltonian constraint: since the generator is a four-vector living in the $Spin(3,1)$ representation space, it must transform as such under a local Lorentz transformation. We will have more to say on this matter shortly.

Let us now consider the commutator of two Hamiltonian constraints.
It will be useful to split the constraint into two separate pieces $C_{H}=C_{H_{0}}+C_{H_{\Lambda}}$ where
\beqa
C_{H_{0}}(t)&=&\frac{1}{k}\int_{\Sigma}-[t, e]\,P_{\star}\,R \\
C_{H_{\Lambda}}(t)&=&\frac{1}{k}\int_{\Sigma}\ts{\frac{\Lambda}{3}}\,[t, e]\,\star e\, e\,.
\eeqa
The commutator we wish to evaluate now becomes,
\beqa
\{C_{H}(t_{1}),C_{H}(t_{2})\}&=&\{C_{H_{0}}(t_{1}),C_{H_{0}}(t_{2})\}
+\{C_{H_{\Lambda}}(t_{1}),C_{H_{\Lambda}}(t_{2})\}\nn\\
& & +\{C_{H_{0}}(t_{1}),C_{H_{\Lambda}}(t_{2})\}
+\{C_{H_{\Lambda}}(t_{1}),C_{H_{0}}(t_{2})\}\,.\nn\\
& &
\eeqa
Since $C_{H_{\Lambda}}$ does not contain $\omega$, clearly we have
\beq
\{C_{H_{\Lambda}}(t_{1}), C_{H_{\Lambda}}(t_{2})\}=0\,.
\eeq
Computing cross-terms we have
\beqa
\{C_{H_{0}}(t_{1}), C_{H_{\Lambda}}(t_{2})\}
&+&\{C_{H_{\Lambda}}(t_{1}), C_{H_{0}}(t_{2})\}\nn\\
&=& \frac{1}{k}\int_{\Sigma}{\ts\frac{\Lambda}{2}}\star [t_{2}, e]\,D[t_{1}, e]
-\frac{1}{k}\int_{\Sigma}{\ts\frac{\Lambda}{2}}\star [t_{1}, e]\,D[t_{2}, e] \nn\\
&=& 0\, .
\eeqa
Thus, we see that the commutator reduces to
\beq
\{C_{H}(t_{1}), C_{H}(t_{2})\}=\{C_{H_{0}}(t_{1}), C_{H_{0}}(t_{2})\}\,.
\eeq
Proceeding, we take the gradient of $C_{H_{0}}$ which yields, upon identification of components:
\beqa
(e\, P_{\star} X^{(\omega)}_{1} 
+P_{\star} X^{(\omega)}_{1}  e)&=& [t_{1}, P_{\star}R] \nn\\
P_{\star}(X^{(e)}_{1}\, e +e\, X^{(e)}_{1}) &=& P_{\star}\,D[t_{1},e]\,.
\eeqa
The symplectic form contracted onto the canonical vector fields takes the general form
\beqa
\bm{\Omega(\bar{X}_{C_{H_{0}}(t_{1})},\bar{X}_{C_{H_{0}}(t_{2})})}&=&
\frac{1}{k}\int_{\Sigma} P_{\star}X^{(\omega)}_{1}\,(X^{(e)}_2\,e+e\,X^{(e)}_{2}) \nn\\
& &-P_{\star}X^{(\omega)}_{2}\,(X^{(e)}_{1}\,e+e\,X^{(e)}_{1}) 
\eeqa
Inserting the $X^{(\omega)}$ components first we have
\beq
\frac{1}{k}\int_{\Sigma}[t_{1},P_{\star}R]\,X^{(e)}_2- [t_{2},P_{\star}R]\,X^{(e)}_1
\,. \label{HalfCom}
\eeq
At this point we appear to be stuck. Only the particular combination $P_{\star}(X^{(e)}\, e +e\,X^{(e)})$ of the canonical
vector field associated with $C_{H_{0}}$ are determined from the symplectic form, yet we simply need the components
$X^{(e)}$ to evaluate the above. We could attempt to re-evaluate the expression by inserting the $X^{(e)}$ components
first and we arrive at 
\beqa
\{C_{H_{0}}(t_{1}),C_{H_{0}}(t_{2})\}= \frac{1}{k}\int_{\Sigma}P_{\star}X^{(\omega)}_1\,[t_{2},T]
-P_{\star}X^{(\omega)}_2 \,[t_{1},T] \,.
\eeqa
We see we are stuck with the same problem---we need the components of the canonical vector field
$X^{(\omega)}$, but only the combination, $e\, P_{\star} X^{(\omega)} 
+P_{\star} X^{(\omega)} \, e$, is given by the symplectic form. The root of the problem is that the constraint,
$C_{H_{0}}$, contains only one factor of $e$, whereas the symplectic structure is quadratic in $e$. We were able to
evaluate the other commutators because at least one of the constraints in the commutator was quadratic or more in $e$.
In the next section we will show that the commutator can be evaluated by a method of partial inverses.

\subsection{Resolving the commutator $\{C_{H}(t_{1}),C_{H}(t_{2})\}$}
In the previous section we reached an impasse in evaluating the commutator of two Hamiltonian constraints. The problem
essentially boiled down to the constraint being linear as opposed to quadratic in the tetrad. Here we will show that
we can, in fact, resolve this problem by considering the partial inverse of the dynamical field $e$, which we detail
in appendix \ref{Inverses}. The main point is that our initial assumption of invertibility of the tetrad $\vep$
implies conditions on the three dimensional dynamical variable $e$ and the Lagrange multiplier $t$. These conditions
are not dimension reducing, just as requiring invertibility of an $n\times n$ matrix does not reduce the dimension of
the space: $GL(n)$ still has dimension $n^{2}$. However, the condition does restrict the possible configurations of $e$ and $t$. In particular, invertibility of the tetrad implies the restriction that only those configurations, $(e^{I},t^{J})$, are allowed that admit ``inverse" fields
$\bar{\theta}_{I}=\theta^{a}_{I}\,\frac{\partial}{\partial x^{a}}$ and $\theta_{J}$ such that
\beqa
e^{I}_{a}\,\theta^{b}_{I}&=&\delta^{a}_{b}\nn\\
e^{I}_{a}\,\theta^{a}_{J}&=&\delta^{I}_{J}-t^{I}\theta_{J}\nn\\
e^{I}_{a}\,\theta_{I}&=&0\nn\\
\theta^{a}_{I}t^{I}&=&0\nn\\
t^{I}\theta_{I}&=&1\ . \label{InverseRelations}
\eeqa
Given a tetrad on, $\vep^{I}_{\mu}$, on $M$ such that $e^{I}_{a}\equiv \vep^{I}_{a}$ and $t^{J}\equiv \vep^{J}_{0}$, and
its (unique) inverse $\vartheta^{\beta}_{J}$, we can identify the inverse fields by $\theta^{a}_{I}\equiv\vartheta^{a}_{I}$
and $\theta_{J}\equiv\vartheta^{0}_{J}$. The above relations are simply the inverse formulas
$\vep^{I}_{\mu}\vartheta^{\nu}_{I}=\delta^{\nu}_{\mu}$ and $\vep^{I}_{\mu}\vartheta^{\mu}_{J}=\delta^{I}_{J}$. We
note, that the fields $\bar{\theta}_{I}=\bar{\theta}_{I}(e,t)$ and $\theta_{J}=\theta_{J}(e,t)$ are explicit functions of
$e$ and $t$. More details on the inverse relations and the relation these variables to the lapse and shift can
be found in appendix \ref{Inverses}. 

We recall that the bare components of the Hamiltonian vector field associated with a function $f$ are
not determined by Hamilton's equations in this formalism, but only the combinations
\beqa
e\, P_{\star} X^{(\omega)}_f 
+P_{\star} X^{(\omega)}_f  e &=& k\,\frac{\delta f}{\delta e} \nn\\
P_{\star}(X^{(e)}_f e +e\, X^{(e)}_f e) &=& -k\,\frac{\delta f}{\delta \omega}\ .
\eeqa
Let us focus on the second of these two equations. Assuming the tetrad is invertible, we can invert this equation to solve for $X^{(e)}$ (as a function of $t$ and $\lambda$).
To this end, define $\chi\equiv -k\,P^{-1}_{\star}\frac{\delta f}{\delta\omega}$. The expression above then reduces to 
\beq
X^{(e)}_f\, e +e\, X^{(e)}_f=\chi\,.
\eeq
From the properties of the inverse fields, $\bar{\theta}_{I}(e,t)$ and $\theta_{J}(e,t)$, we can solve for $X^{(e)}_f$
as a function of $e$ and $t$ as follows: 
\beq
(X^{(e)}_f)^{I}=2\chi^{MI}(\bar{\theta}_{M}, \ )-t^{I}\theta_{N}\chi^{MN}(\bar{\theta}_{M},\ )-
{\ts\frac{1}{2}}e^{I}\,\chi^{MN}(\bar{\theta}_{M},\bar{\theta}_{N})\,.\label{deltae}
\eeq
This expression allows for a general definition of the commutator of two functions, $f$ and $g$. 

We will now use this method to calculate the commutator of two Hamiltonian constraints. We recall that the commutator
$\{C_{H}(t_{1}), C_{H}(t_{2})\}$ reduces to $\{C_{H_{0}}(t_{1}), C_{H_{0}}(t_{2})\}$.
Hamilton's equation $\bm{\Omega}(\bm{\bar{X}_{C_{H_0}(t)}},\ )=\dl C_{H_{0}}(t)$ yields the expressions
\beqa
e\, P_{\star} X^{(\omega)}
+P_{\star} X^{(\omega)}\, e &=& [t, P_{\star}R] \nn\\
X^{(e)} e +e\, X^{(e)} &=& D[t,e]\,.
\eeqa
From the procedure described above, we can solve the second equation to yield
\beq
X^{(e)} =Dt-{\ts\frac{1}{2}}t\,T^{I}(\bar{\theta}_{I})\,.
\eeq
Thus, the commutator reduces to
\beqa
\{C_{H}(t_{1}),C_{H}(t_{2})\}&=&\bm{\Omega(\bar{X}_{C_H (t_1)},\bar{X}_{C_H (t_2)})}\nn\\
&=&\frac{1}{k}\int_{\Sigma}[t_{1},P_{\star}R]\,\left(Dt_{2}-{\ts\frac{1}{2}}t_{2}T^{M}(\bar{\theta}^{(2)}_{M})\right)\ -\
\{1\leftrightarrow 2\}\nn\\
&=&\frac{1}{k}\int_{\Sigma}{\ts\frac{1}{2}}[t_{1},t_{2}]\,P_{\star}R\
T^{M}(\bar{\theta}^{(1)}_{M}+\bar{\theta}^{(2)}_{M}) \label{HHfinal}
\eeqa
where we have used the notation $\bar{\theta}^{(1)}_{I}=\bar{\theta}_{I}(e,t_{1})$ and
$\bar{\theta}^{(2)}_{I}=\bar{\theta}_{I}(e,t_{2})$.

\subsection{Properties of the commutator}
There are several interesting properties that we can derive from the above. First we note that final result for the commutator is explicitly dependent on the torsion via $T^{I}(\bar{\theta}_{I},\ )$. Let us consider the solution to the Gauss constraint, here
written in unsmeared form:
\beq
D(e^{I}\,e^{J})=T^{I}\,e^{J}-e^{I}\,T^{J}\approx 0\,.
\eeq
Some simple algebra shows that we can (partially) invert this expression to give $\theta_{I}\,T^{I}\approx 0$ and
$T^{J}(\bar{\theta}_{J},\ )\approx 0$. We note that this must hold for any $\theta_{I}(e,t)$ and
$\bar{\theta}_{I}(e,t)$ that satisfies (\ref{InverseRelations}).
This in turn implies that the
commutator of two Hamiltonian constraints is weakly vanishing:
\beq
\{C_{H}(t_{1}), C_{H}(t_{2})\}\approx 0 \,.
\eeq
Thus, the algebra closes without the need for introduction of new constraints. All that we have assumed is the existence of canonical vector fields associated with the total Hamiltonian. In
retrospect this was a foregone conclusion. After all, the canonical variables are simply the pull-back of the
dynamical Lagrangian variables to $\Sigma$, and Hamilton's equations together with the vanishing of the constraints are themselves precisely the Einstein equations
pulled-back to $\Sigma$ as seen by equations (\ref{HEOMa}) and (\ref{HEOMb}). The symplectic evolution of the system simply
gives us the remaining components of the Einstein-Cartan equations. Thus, the question of whether the
constraint algebra closes weakly is equivalent to the question: {\it are Einstein's equations self-consistent?} The
answer is, of course, {\it yes!} Phrased another way, suppose we have a set of initial data 
$\omega_{t_{0}}$ and $e_{t_{0}}$ on the initial Cauchy surface $\Sigma_{t_{0}}$. If the data set is a good data set, it
will solve the constraint equations---in other words it will solve Einstein's equations pulled-back to
$\Sigma_{t_{0}}$. The symplectic evolution simply enforces the remaining equations of motion, but for our purposes it
also serves to evolve $\omega_{t_{0}}$ and $e_{t_{0}}$ on $\Sigma_{t_{0}}$ to $\omega_{t_{0}+\Delta t}$ 
and $e_{t_{0}+\Delta t}$ on the new Cauchy surface $\Sigma_{t_{0}+\Delta t}$. Now the questions is {\it does the new
data satisfy Einstein's equations pulled back to the new Cauchy slice?} If it does then it will satisfy the
constraints on $\Sigma_{t_{0}+\Delta t}$. If it doesn't then this will be reflected in the non-closure of the
constraints, which would indicate that the evolution generated by the given constraints and symplectic
structure pulls the initial data off the constraint submanifold. This, in turn, would indicate a need for more
constraints. But, since two of the constraints plus the evolution equation are {\it precisely} the full set of Einstein's
equations, and Einstein's equations are self-consistent (barring the emergence of singularities
where various physical quantities become singular), this cannot happen.
So the constraint algebra {\it must} close. 

Another striking property emerges from the constraint algebra. We first notice that apart from the
commutator of two Hamiltonian constraints, the algebra is precisely the (A)dS/Poincar\'{e} algebra. Thus, let us focus on the
expression (\ref{HHfinal}), repeated here:
\beq
\{C_{H}(t_{1}),C_{H}(t_{2})\}=\frac{1}{k}\int_{\Sigma}{\ts\frac{1}{2}}[t_{1},t_{2}]\,\PS R\
T^{M}(\bar{\theta}^{(1)}_{M}+\bar{\theta}^{(2)}_{M})\,.
\eeq
We wish to evaluate the right hand side on an arbitrary solution to the equations of motion in order to see the
relation with the (A)dS/Poincar\'{e} algebra. Thus, consider a four-dimensional field configuration $\varpi$ and $\vep$ that 
satisfies the Einstein-Cartan equation (\ref{EOM1}), but not necessarily the torsional constraint (\ref{EOM2}) (otherwise
the commutator would trivially vanish on this solution). Using this equation of motion, we derive the following identity on this {\it partial} solution subspace:
\beq
Tr\left([t_1,t_2]\PS R \right)=Tr\left(e\,e\,\PS R (\bar{t}_1,\bar{t}_2) \right)\,.
\eeq
Inserting this expression into the commutator we have
\beqa
\{C_{H}(t_{1}),C_{H}(t_{2})\}& \approx &\frac{1}{k}\int_{\Sigma}{\ts\frac{1}{2}}\,e\,e\,\PS 
R(\bar{t}_1,\bar{t}_2)\,
T^{M}(\bar{\theta}^{(1)}_{M}+\bar{\theta}^{(2)}_{M})\nn\\
&=&\frac{1}{k}\int_\Sigma D (R(\bar{t}_1,\bar{t}_2))\, \PS e\,e\nn\\
&=& C_G(-R(\bar{t}_1,\bar{t}_2))\,.
\eeqa
Let us now consider the general solution to the equation of motion (\ref{EOM1}) given by
\beq
R_\varpi=\ts{\frac{\Lambda}{3}}\,\vep\,\vep +\mathcal{C}
\eeq
where $\mathcal{C}$ is the conformal Weyl tensor. Inserting this into the expression above, we arrive at a particularly simple form for the commutator:
\beqa
\{C_{H}(t_{1}),C_{H}(t_{2})\}&\approx&\frac{1}{k}\int_{\Sigma}D\left(\ts{\frac{\Lambda}{3}}[t_{1},t_{2}]+\mathcal{C}(\bar{t}_{1},\bar{t}_{2})\right)\star e\,e
\nn\\
 &=& -{\ts\frac{\Lambda}{3}}C_{G}([t_{1},t_{2}]) -C_{G}(\mathcal{C}(\bar{t}_{1},\bar{t}_{2}))\,. \label{ReducedHH}
\eeqa
Thus, as predicted, the constraint algebra does reduce down to a deformation of the de Sitter, anti-de Sitter, or Poincar\'{e} algebra depending on the value of the cosmological constant. Furthermore, we
see the local degrees of freedom of general relativity emerging from the constraint algebra itself in the form of the
conformal Weyl tensor, which plays the role of the (not necessarily small) deformation parameter. Evaluated on the homogenous solution, where the Weyl tensor vanishes the constraint algebra reduces precisely to the (A)dS/Poincar\'{e} algebra as expected since diffeomorphisms on maximally symmetric spacetimes are equivalent to gauge transformations.

\section{A consistency check}
In a previous paper \cite{Randono:CanonicalLagrangian} we developed an alternative method for computing the constraint algebra restricted to the
constraint submanifold $\bar{\Gamma}_{H}$. This method focused on the Lagrangian one-form, $\bm{\theta}$, given
explicitly by (\ref{Lagrangian1form}). There it was shown that, with some caveats that we will make more explicit shortly, we can identify
the Poisson bracket with the following expression\footnote{We have made a slight change of convention from version one of \cite{Randono:CanonicalLagrangian} in order to match prevailing conventions and avoid potentially confusing minus signs in the commutator. There the commutator was defined as $\{f,g\}\equiv \bm{\Omega(\bar{X}_g,\bar{X}_f)}=\bm{\mathcal{L}_{\bar{X}_f}}g=-\bm{\mathcal{L}_{\bar{X}_g}}f$. In this paper, the commutator is defined with a minus sign: $\{f,g\}\equiv \bm{\Omega(\bar{X}_f,\bar{X}_g)}=-\bm{\mathcal{L}_{\bar{X}_f}}g=\bm{\mathcal{L}_{\bar{X}_g}}f$.}:
\beq
\{C_{tot}(t_{1},\lambda_{1})\,,\,C_{tot}(t_{2},\lambda_{2})\}\approx
\bm{\theta([\bar{t}_{1},\bar{t}_{2}])}
\eeq
where
\beq
\bm{[\bar{t}_{1},\bar{t}_{2}]}=\int_{M}\mathcal{L}_{[\bar{t}_{1},\bar{t}_{2}]}\varpi\,\frac{\dl}{\dl\varpi}
+\mathcal{L}_{[\bar{t}_{1},\bar{t}_{2}]}\vep\,\frac{\dl}{\dl\vep}\,.
\eeq
Thus, we can use this expression as a consistency check on our previous derivation. In deriving the expression above, all exterior derivatives are derivatives on the full Lagrangian phase space. Thus, the Lagrange multipliers $\lambda=-\varpi(\bar{t})$ and $t=\vep(\bar{t})$ are varied in the above expression, effectively making them dynamical variables. To make contact with the standard Hamiltonian formalism we have to take this into account. The only difference between the expression above and the standard commutator are terms involving the time derivatives of the Lagrange multipliers, as follows (here the notation $F \mid_{t, \lambda}$ means that the function $F$ is computed while holding $t$ and $\lambda$ fixed when derivatives are involved) :
\beqa
\bm{\mathcal{L}_{\bar{t}_1}}C_{tot}(t_2,\lambda_2)\big|_{t_2,\lambda_2}&=&\bm{\theta([\bar{t}_1,\bar{t}_2])}+C_H(t^{[12]})
+C_G(\lambda^{[12]})\nn\\& &+i_{\bm{\bar{t}_2}}\left(\bm{\Omega(\bar{t}_1,\ )}-\dl C_{tot}(t_1,\lambda_1)\big|_{t_1,\lambda_1}\right)
\eeqa
where $t^{[12]}\equiv \mathcal{L}_{\bar{t}_1}t_2-\mathcal{L}_{\bar{t}_2}t_1$ and $\lambda^{[12]}\equiv \mathcal{L}_{\bar{t}_1}\lambda_2-\mathcal{L}_{\bar{t}_2}\lambda_1$. Thus, on the submanifold $\Gamma^{(2)}_H$ where canonical vector fields associated with the total Hamiltonian exist, the second line in the equation above vanishes and we have
\beqa
\{C_{tot}(t_1,\lambda_1),C_{tot}(t_2,\lambda_2)\}\big|_{t_{1,2},\lambda_{1,2}} \stackrel{\Gamma^{(2)}_H}{\approx} \bm{\theta([\bar{t}_1,\bar{t}_2])}+C_H(t^{[12]})
+C_G(\lambda^{[12]})\,.
\eeqa
Since the Lagrangian one-form is itself used to construct the Hamiltonian, the above expression is easy to compute using the identity
\beq
\bm{\theta([\bar{t}_{1},\bar{t}_{2}])}=C_{H}(-\vep([\bar{t}_{1},\bar{t}_{2}]))+C_{G}(\varpi([\bar{t}_{1},\bar{t}_{2}]))\,.
\eeq
A straightforward calculation yields the identities
\beqa
-\vep([\bar{t}_{1},\bar{t}_{2}])&=&T(\bar{t}_{1},\bar{t}_{2})+[\lambda_{1},t_{2}]-[\lambda_{2},
t_{1}]-t^{[12]} \nn\\
\varpi([\bar{t}_{1},\bar{t}_{2}])&=&-R(\bar{t}_{1},\bar{t}_{2})+[\lambda_{1},\lambda_{2}]
-\lambda^{[12]}\,,
\eeqa
the total expression for the commutator becomes
\beqa
\{C_{tot}(t_{1},\lambda_{1})\,,\,C_{tot}(t_{2},\lambda_{2})\}\big|_{t_{1,2},\lambda_{1,2}}&=&C_{H}(T(\bar{t}_{1},\bar{t}_{2})+[\lambda_{1},t_{2}]-[\lambda_{2},t_{1}] )\nn\\
& & +C_{G}(-R_{\varpi}(\bar{t}_{1},\bar{t}_{2})+[\lambda_{1},\lambda_{2}])\,.
\eeqa

On $\bar{\Gamma}_{H}$ Hamilton's equations of motion hold, which we have shown are identical to the Einstein equations
pulled-back to the three manifold. Thus, on this submanifold, the general solution is $T=0$ and
$R_{\varpi}=\frac{\Lambda}{3}\vep\,\vep+\mathcal{C}$, where $\mathcal{C}$ is the conformal Weyl tensor. Thus, on $\bar{\Gamma}_{H}$ we have
\beqa
\{C_{tot}(t_{1},\lambda_{1}),C_{tot}(t_{2},\lambda_{2})\}\big|_{t_{1,2},\lambda_{1,2}}=C_{H}([\lambda_{1},t_{2}]-[\lambda_{2},t_{1}])\nn\\
+ C_{G}([\lambda_{1},\lambda_{2}]-\ts{\frac{\Lambda}{3}}[t_{1},t_{2}]-\mathcal{C}(\bar{t}_{1},\bar{t}_{2}))\,.
\eeqa
This is consistent with our previous expressions for the
commutators (\ref{Commutator1}) and (\ref{ReducedHH}). Once again, we see that the Poisson algebra of $C_H(t)$ and $C_G(\lambda)$ is a deformation of the (A)dS or Poincar\'{e} algebra with the deformation parameter begin the conformal Weyl tensor.

\section{The relation between diffeomorphisms and the vectorial Hamiltonian constraint}
The theory expounded in the previous sections is diffeomorphism invariant since it is simply general relativity written in Hamiltonian language.
However, diffeomorphism symmetry enters into the construction in a subtle way. The natural explanation
would seem to be that the three--dimensional diffeomorphisms have been absorbed into the Hamiltonian constraint
explaining the extra degrees of freedom of the constraint. This is partially true---the information contained in a
diffeomorphism gauge transformation is contained in the Hamiltonian constraint. However, the information is embedded
in such a way that one cannot extract a three-dimensional diffeomorphism from the Hamiltonian constraint without
breaking the gauge. In this sense, the vectorial nature of the Hamiltonian constraint is not simply a convenient
repackaging of the ordinary diffeomorphism and scalar constraints. This, in part, explains the discrepancies in the
constraint algebra from other covariant approaches \cite{Alexandrov:Covariant}.

In order to understand the nature of the diffeomorphism symmetry in this theory we first need to set up some
preliminaries. It will be useful to distinguish the fiber preserving diffeomorphisms from the gauge
covariant diffeomorphisms. We will restrict ourselves to one-parameter diffeomorphisms connected to the identity. By
``fiber-preserving" we mean the ordinary action of the diffeomorphisms on Lie-algebra valued forms defined infinitesimally by:
\beq
\alpha\rightarrow \phi_{\bar{N}}\alpha=\alpha+\mathcal{L}_{\bar{N}}\alpha 
\eeq
where $\mathcal{L}_{\bar{N}}\alpha=i_{\bar{N}}d\alpha+d(i_{\bar{N}}\alpha)$ is the ordinary Lie derivative with respect to a smooth vector field $\bar{N}\in T\Sigma$. We call
this transformation fiber preserving because it is not gauge covariant---it fixes a basis in the fiber. To circumvent
this, we define the gauge covariant Lie derivative, $\mathfrak{L}_{\bar{N}}\alpha$ such that under an infinitesimal $Spin(3,1)$
gauge transformation we have:
\beq
\mathfrak{L}_{\bar{N}}[\lambda,\alpha]=[\lambda,\mathfrak{L}_{\bar{N}}\alpha]. \label{GCLie}
\eeq
On Lie algebra valued forms, the gauge covariant Lie derivative is then:
\beq
\mathfrak{L}_{\bar{N}}\alpha=i_{\bar{N}}D\alpha+D(i_{\bar{N}}\alpha)
\eeq
and the action on a connection coefficient is defined so that $\mathfrak{L}_{\bar{N}}R=D\mathfrak{L}_{\bar{N}}\omega$
from which we have
\beq
\mathfrak{L}_{\bar{N}}\omega=R(\bar{N}).
\eeq
Following the general Hamiltonian program, we can implement the action of these transformations on the phase space using the
Hamiltonian vector fields of some functionals, which we denote $\mathcal{D}(\bar{N})$ and $\mathfrak{D}(\bar{N})$ for
the fiber preserving and gauge covariant diffeomorphisms respectively. These functionals are:
\beqa
\mathcal{D}(\bar{N})&=&\int_{\Sigma}\mathcal{L}_{\bar{N}}\omega P_{\star}e\,e\\
\mathfrak{D}(\bar{N})&=&\int_{\Sigma}\mathfrak{L}_{\bar{N}}\omega P_{\star}e\,e\,.
\eeqa
The constraint algebra of
$\mathcal{D}(\bar{N})$ is that of the ordinary diffeomorphisms:
\beqa
\{ \mathcal{D}(\bar{N}_{1}), \mathcal{D}(\bar{N}_{2})\}&=&\mathcal{D}([\bar{N}_{1},\bar{N}_{2}])\nn\\
\{\mathcal{D}(\bar{N}), C_{G}(\lambda)\}&=&C_{G}(\mathcal{L}_{\bar{N}}\lambda)\nn\\
\{\mathcal{D}(\bar{N}), C_{H}(\eta)\}&=& C_{H}(\mathcal{L}_{\bar{N}}\eta)\,.
\eeqa
On the other hand, $\mathfrak{D}(\bar{N})$ obeys a different algebra, most notably, as a reflection of (\ref{GCLie}) we
have:
\beq
\{\mathfrak{D}(\bar{N}), C_{G}(\lambda)\}=0\,. \label{GCLieConstraint}
\eeq

Now, the constraint and $\mathfrak{D}(\bar{N})$ vanishes on the submanifold defined by $C_{H}\approx 0$ and 
$\mathcal{D}(\bar{N})$ vanishes on the submanifold defined by $C_{H}\approx 0$ and $C_{G}\approx 0$. Noting the
following identities
\beqa
\mathfrak{D}(\bar{N}) &=& C_{H}(e(\bar{N}))\nn\\
\mathcal{D}(\bar{N}) &=& \mathfrak{D}(\bar{N})+C_{G}(\omega(\bar{N}))\,, \label{DequalsH}
\eeqa
it is clear that the physical constraint submanifold has both fiber preserving and gauge
covariant diffeomorphism symmetry. On the other hand, we note that since the generators in $C_{H}(e(\bar{N}))$ 
and $C_{G}(\omega(\bar{N}))$ are field dependent, $\mathfrak{D}(\bar{N})$ and $\mathcal{D}(\bar{N})$ obey very
different Poisson algebras, and generate very different symmetries than $C_{H}(t)$ and $C_{G}(\lambda)$.

We now come to the crucial point. From the identity (\ref{DequalsH}), it is natural to try to divide the Hamiltonian
constraint into $C_{H}(t)\stackrel{?}{=}C_{H}(t_{\perp})+C_{H}(t_{\|})$ where $t_{\|}\cdot t_{\perp}=0$, and $t_{\|}=e(\bar{N})$ where $\bar{N}$ is the shift. The splitting should be such that $C_{H}(t_{\perp})$ generates the ordinary
``lapse" transformations of most canonical approaches, and $C_{H}(t_{\|})$ generates diffeomorphisms. From the
above discussion, we would then have $C_{H}(t_{\|})=C_{H}(e(\bar{N}))=\mathfrak{D}(\bar{N})$. If this could be done,
then the diffeomorphisms would not be an independent symmetry from the Hamiltonian constraint, and this approach would
simply be a convenient way of repackaging the ordinary scalar and diffeomorphism constraints. Suppose then, that
we could do this. Then consider the evolution of the Hamiltonian constraint along the Hamiltonian vector field,
$\bm{\bar{X}}_{\lambda}$, associated with $C_{G}(\lambda)$:
\beq
-\bm{\mathcal{L}_{\bar{X}_\lambda}}C_{H}(t)=\{C_{G}(\lambda),C_{H}(t)\}=C_{H}([\lambda,t]).
\eeq
On the other hand, from (\ref{GCLieConstraint}) we have
\beqa
-\bm{\mathcal{L}_{\bar{X}_\lambda}}(C_{H}(t_{\perp})+\mathfrak{D}(\bar{N}))
&=&\{C_{G}(\lambda),C_{H}(t_{\perp})+\mathfrak{D}(\bar{N})\} \\
&=& C_{H}([\lambda,t_{\perp}])\,.
\eeqa
It is clear from the last equality that the splitting of the time evolution vector field as such is not a covariantly defined operation. Furthermore, from the last equality it is clear that the split is not even Lorentz {\it invariant} unless we restrict ourselves to a subgroup of the local Lorentz group that preserves the vector $t_\perp$, which is equivalent to a breaking of the local gauge group to its rotation subgroup. In this case, the Gauss constraint commutes with the new Hamiltonian constraint, implying that it is truly a scalar constraint. Indeed, rewriting $t_\perp =\alpha\,n$ where $\alpha$ is the lapse, and $n$ is the normal to $\Sigma$ lifted to the fiber, and fixing once and for all the direction of the normal so that $n^I=(1,0,0,0)$ (we can do this once we have fixed the gauge), we obtain the ordinary scalar constraint in the time gauge (subject to the imposition of second class constraints that emerge upon computing the constraint algebra). The remaining piece, $\mathfrak{D}(\bar{N})$, naturally becomes the vector constraint in the time gauge. Thus, in this respect, the vector Hamiltonian of our theory in some sense ``contains" both the scalar and the vector constraints of the standard formalism, but one must be careful not to take this interpretation too literally since they are contained in such a way that any splitting of the Hamiltonian constraint into scalar plus vector necessarily breaks the gauge.

\section{The physical degrees of freedom}
In the preceeding sections we have given a Hamiltonian theory whose constraints and Hamilton equations of motion are
general relativity, and whose constraint algebra is first class. On the other hand, a simple counting argument
suggests a paradox: the total physical degrees of freedom don't seem to yield the accepted two degrees of freedom of
general relativity. To see this, we need to count the kinematical degrees of freedom minus the degrees of freedom of
the constraint algebra. Typically, to count the kinematical degrees one counts the local degrees of freedom of either
the momentum or the position variable (just the configuration space), and then proceeds to subtract out the constraints. In our case, the symplectic
structure does not polarize the phase space in any obvious way, so we simply count the total degrees on the full constrained phase space and once the final tally is obtained we divide by two. For general relativity, we expect the final tally for the phase space degrees of freedom to
be $4$, or dividing by two, we should obtain standard result of two ``configuration space" degrees of freedom. At
the kinematical level (prior to the implementation of any constraints) the total degrees of freedom in the Hamiltonian phase space is given by
\beqa
DOF_{kin}&=& DOF(e)+DOF(\omega)=3\times 4+3\times 6=30\,.
\eeqa
The physical degrees of freedom are then obtained by subtracting out the constraints and the spurious gauge degrees of
freedom. Let us first recall then how the gauge degrees of freedom are determined. We begin with the kinematical
phase space $\bm{\Gamma}$ where the kinematical dynamical variables form a good set of coordinates and the symplectic
structure is typically non-degenerate. We define the constraint submanifold $\bm{\widetilde{\Gamma}}$ to be the
submanifold on which a set of first class constraints vanish weakly: $C_{\{i\}}\approx 0$. Let
$\phi:\bm{\widetilde{\Gamma}}\rightarrow \bm{\Gamma}$ be the embedding of $\bm{\widetilde{\Gamma}}$ in
$\bm{\Gamma}$. The pullback of the symplectic structure, $\bm{\widetilde{\Omega}}\equiv \phi^{*}\bm{\Omega}$ is now
degenerate. To see this, recall the definition of the Hamiltonian vector field $\bm{\bar{X}}_{\{i\}}$ associated with
$C_{\{i\}}$ is $\bm{\Omega}(\bm{\bar{X}}_{\{i\}},\ )=\dl C_{\{i\}}$. Since $C_{\{i\}}$ is constant on
$\bm{\widetilde\Gamma}$ (in fact zero), the pull-back of the gradient of the constraint is zero: $\phi^{*}\dl
C_{\{i\}}=0$. Since the constraints are first class, by definition the Hamiltonian vector fields
$\bm{\bar{X}}_{\{i\}}$ are parallel to the constraint submanifold. That is,
they can be identified with the push-forward of some vector, $\bm{\bar{Y}}_{\{i\}}$ in $T\,\bm{\tilde{\Gamma}}$:
\beq
\bm{\bar{X}}_{\{i\}}\equiv \phi_{*}\bm{\bar{Y}}_{\{i\}}.
\eeq
In total, the pull-back of the Hamiltonian condition gives:
\beqa
\phi^{*}\bm{\Omega}(\bm{\bar{X}}_{\{i\}},\ )&=&\phi^{*}\dl C_{\{i\}}=0\nn\\
&\rightarrow & \bm{\widetilde{\Omega}}(\bm{\bar{Y}}_{\{i\}},\ )=0
\eeqa
We see from the second line above that the symplectic structure pulled back to $\bm{\widetilde{\Gamma}}$ is
degenerate, and the kernel is spanned by the vector fields $\bm{\bar{Y}}_{\{i\}}$. These directions are precisely the
gauge directions of the constraint submanifold---evolving a point on the phase space in these directions is equivalent
to a gauge transformation and will not change the physics of the problem at hand.

From the above discussion, it is clear that the role of the constraints is two-fold. First, the constraints serve to
define the physical submanifold thereby reducing the (local) dimension of the physical degrees of freedom to the (local)
dimension of the constraint submanifold. Second, the constraints define gauge directions on the constraint
submanifold, which further reduce the dimension of the true, physical degrees of freedom. Thus, each constraint reduces
the kinematical phase space degrees of freedom by $2\times DOF(C_{\{i\}})$. In our case, we have
\beqa
DOF_{phys}&\stackrel{?}{=}&DOF(e)+DOF(\omega)-(2\times DOF(C_{H})+2\times DOF(C_{G}))\nn\\
&=& 3\times 4+3\times 6 - (2\times 4 +2\times 6) \nn\\
&=& 10\,.
\eeqa
We now have a paradox. The true physical degrees of freedom in the phase space should be $4$ (or two configuration space degrees of freedom), but we
appear to have $10$ degrees of freedom. Where are the six extraneous degrees of freedom coming from? 

The answer is that we have not taken into account all the gauge degrees of freedom. If we generalize the concept of
``gauge", to include all degenerate directions of the physical symplectic form, then we allow for gauge degrees of
freedom that are not associated with any constraint at all. This would occur when the original symplectic form
$\bm{\Omega}$ is itself degenerate. In this case, there would be some vector fields $\bm{\bar{Z}}$ such that
$\bm{\Omega}(\bm{\bar{Z}},\ )=0$. Such vector fields would generate transformations, which could equally well be
considered gauge transformations since they suggest an indeterminancy in the definition of the Hamiltonian vector
fields.

 In fact, this is precisely the situation that we have in our model. We recall that the (pre)symplectic
form, $\bm{\Omega}=\frac{1}{k}\int_{\Sigma}P_{\star}\dl\omega\wedge\dl (e\,e)$, is degenerate. 
Given some vector field $\bm{\bar{Z}}$ in $T\Gamma_H$ that is in the kernel of
$\bm{\Omega}$ so that $\bm{\Omega}(\bm{\bar{Z}},\ )=0$, write the vector fields in component notation
\beq
\bm{\bar{Z}}=\int_{\Sigma}Z^{(e)}\frac{\dl}{\dl e}+Z^{(\omega)}\frac{\dl}{\dl \omega}
\eeq
where $Z^{(e)}=\frac{1}{2}\gamma_{I}\,(Z^{(e)})^{I}_{a}\,dx^{a}$ and
$Z^{(\omega)}=\frac{1}{4}\gamma_{I}\gamma_{J}\,(Z^{(\omega)})^{[IJ]}_{a}\,dx^{a}$. As shown in appendix \ref{AppendixB}, using the inverse properties of $e$, the degeneracy condition yields $Z^{(e)}=0$. An arbitrary $({Z^{(\omega)}})^{KL}_{a}$ has $3\times 6=18$
degrees of freedom, and the degeneracy condition reduces the degrees of freedom of $Z^{(\omega)}$ by twelve, leaving $18-12=6$ degrees of freedom remaining. Thus, we conclude that $\bm{\bar{Z}}$ generates a (pre)symplecto-morphism with $6$ local degrees of freedom.

We can now return to our counting argument. Including the degrees of
freedom in the kernel of $\bm{\Omega}$ in the gauge degrees of freedom we have:
\beqa
DOF_{phys}&=& DOF_{kin}- (2\times DOF(C_{H})+2\times DOF(C_{G})+DOF(ker_{\bm{\Omega}}))\nn\\
&=& 3\times 4 +3\times 6 -(2\times 4+2\times 6 +6) \nn\\
&=& 4\,.
\eeqa
Dividing this result in half we get the standard two local ``configuration space" degrees of freedom of general relativity. From this analysis it is clear that the degeneracy of $\bOmega$ plays an essential role in the Hamiltonian evolution of the system.

Heuristically, one might be tempted to think of this mechanism as follows: suppose we had a more general theory that described general
relativity only when the physical degrees of freedom were constrained by some additional first class constraint
$C'\approx 0$
in addition to the ordinary dynamical constraints, $C_{\{i\}}$, of the theory. Suppose that the kinematical symplectic
structure $\bm{\Omega'}$ were non-degenerate on the larger phase space $\bm{\Gamma'}$.
Let us further suppose that the constraint $C'$ were such that the constrained phase space is precisely the phase space 
$\bm{\Gamma}$ where $e$ and $\omega$ are good coordinates and the symplectic structure structure pulled back to
$\bm{\Gamma}$ is precisely $\bm{\Omega}=\frac{1}{k}\int P_{\star}\dl \omega \wedge \dl (e\,e)$. The push-forwards 
of the degenerate vector fields $\bm{\bar{Z}}$ would then be precisely the Hamiltonian vector fields of the constraint $C'$
since
\beqa
\iota^{*}\bm{\Omega'}(\bm{\bar{Z'}},\ )&=&\iota^{*}\dl C' \nn\\
&\rightarrow & \bm{\Omega}(\bm{\bar{Z}},\ )=0
\eeqa
where $\iota:\bm{\Gamma}\rightarrow \bm{\Gamma'}$ is the embedding and $\bm{\bar{Z}'}=\iota_{*}\bm{\bar{Z}}$. In this
respect, on the constraint submanifold of $C'$, motion along the integral curves of $\bm{\bar{Z}}$ are just ordinary
gauge transformations generated by $C'$. 

This interpretation is tempting, and the language we have used is suggestive of a more fundamental theory where the symplectic form is non-degenerate and whose constrained dynamics yields general relativity. In fact, BF theory is a candidate theory on which the
kinematical symplectic structure is non-degenerate and whose simplicity constraints yield $\bm{\Gamma}$ and the new
symplectic structure $\bm{\Omega}$. However, it is by now well known that the simplicity constraints of BF theory are
not first class \cite{Alexandrov:Covariant,Engle:LQGVertexLong,Engle:LQGVertexShort}. It is not clear whether this is a peculiarity of BF theory or a generic property of the Hamiltonian construction of Einstein-Cartan gravity. The formalism we have developed here ensures that we do not have to
appeal to a more general theory in order to make sense of the system at hand.

\section{Concluding Remarks}
In this paper we have developed a covariant construction of the Hamiltonian framework for canonical Einstein-Cartan gravity that preserves the gauge-like properties of gravity while clearly illustrating the differences. In particular, since a significant portion of the time evolution of the dynamical Lagrangian variables can be identified with an (A)dS/Poincar\'{e} gauge transformation, most of the constraint algebra is simply a reflection of the (A)dS/Poincar\'{e} algebra. This means that the Gauss constraint must be valued in the full $Spin(3,1)$ Lie algebra, and the Hamiltonian must be valued in the vector generators of the algebra. The local degrees of freedom, for example small amplitude gravitational waves in the linearized theory, effectively deform the algebra by the presence of a non-zero conformal Weyl tensor. 

Since the dynamical variables are simply the Lagrangian variables pulled back to the hypersurface, and the full Lorentz group is retained, this method readily suggests generalizations to include fermions. Ultimately, however, the true benefit of this approach may be in the implications for covariant canonical quantum gravity. We hope this method may lead to significant insight into the nature of the second class constraints of the standard covariant approach, as well as insight into the algebraic structure of the constraints that will carry over to the quantum theory.

\begin{appendix}
\section{Summary of inverse relations\label{Inverses}}
In this section of the appendix we summarize formulas relating the tetrad and its inverse to the dynamical
variables on the three-manifold. Throughout this paper we have assumed that the tetrad is invertible. The invertibility
requirement is equivalent to requiring the 4-volume element to be non-zero:
\beq
\widetilde{^{4}V}=\frac{1}{4!}\epsilon_{IJKL}\,\vep^{I}\w\vep^{J}\w\vep^{K}\w\vep^{L}\neq 0\ .
\eeq
This expression pulls-back to a restriction on the 3-volume:
\beq
\widetilde{^3 V}_t\equiv\widetilde{^4V}(\bar{t})=\frac{1}{3!}\epsilon_{IJKL}\,t^{I}\,e^{J}\w e^{K}\w e^{L}\neq 0\ .
\eeq
Invertibility of the tetrad implies that there is a dual frame of vector fields,
$\bar{\vartheta}_{J}=\vartheta^{\nu}_{J}\,\frac{\partial}{\partial x^{\nu}}$, such that 
$\vep^{I}_{\mu} \,\vartheta^{\mu}_{J}=\delta^{I}_{J} $ and $\vep^{I}_{\mu}\,
\vartheta^{\nu}_{I}=\delta^{\nu}_{\mu}$. In terms of the three-dimensional variables,
$e^{I}_{a}\equiv \vep^{I}_{a}$, $t^{J}\equiv \vep^{J}_{0}$, $\theta^{b}_{K}\equiv \vartheta^{b}_{K}$, 
and $\theta_{L}\equiv \vartheta^{0}_{L}$ these two conditions become:
\beqa
e^{I}_{a}\,\theta^{b}_{I}&=&\delta^{a}_{b}\nn\\
e^{I}_{a}\,\theta^{a}_{J}&=&\delta^{I}_{J}-t^{I}\theta_{J}\equiv {P^I}_J \nn\\
e^{I}_{a}\,\theta_{I}&=&0\nn\\
\theta^{a}_{I}t^{I}&=&0\nn\\
t^{I}\theta_{I}&=&1\ .
\eeqa
From the second equation, since $P$ is a projection operator (${P^I}_K{P^K}_J={P^I}_J$), if one takes a vector (or covector), $V^I$,  in the fiber, projects it down to $T\Sigma$, and then attempts to lift it back up to the fiber, one will not obtain the same vector, but only the projection of the vector, ${P^I}_K V^K$, into a three-dimensional subspace orthogonal to $t^I$ and $\theta_J$.

Since, given the matrix $\vep^{I}_{\mu}$ at a point, the inverse matrix $\theta^{\nu}_{J}$, can be uniquely
constructed, the variables $\theta^{a}_{I}$ and $\theta_{J}$ can be viewed as explicit functionals of
$e^{I}_{a}$ and $t^{J}$ whose variations are given by
\beqa
\delta \theta^{a}_{I}(e,t)&=&-\theta^{a}_{K}\theta^{b}_{I}\,\delta e^{K}_{b}-\theta_{I}\theta^{a}_{K}\,\delta t^{K}\\
\delta \theta_{J}(e,t)&=&-\theta_{K}\theta^{b}_{J}\,\delta e^{K}_{b}-\theta_{J}\theta_{K}\,\delta t^{K}\ .
\eeqa
Using just $e^I_a$ one can always project a (co)-vector in the fiber down to the tangent space of
$\Sigma$. However, since $\theta^{a}_{I}=\theta^{a}_{I}(e,t)$, one cannot in general lift a vector from the tangent space to the fiber without first specifying $t^{I}$, which is a gauge choice. Furthermore, even given such a choice, for a generic vector $V_a\in T^*\Sigma$, there is no {\it unique} (co)-vector $V_I$ in the fiber such that $V_I e^I_a=V_a$.

Finally, using the relations $\eta_{IJ}\vep^{I}_{\mu}\vep^{J}_{\nu}={^{4}g_{\mu\nu}}$ and 
$\eta^{IJ}\vartheta^{\mu}_{I}\vartheta^{\nu}_{J}={^{4}g^{\mu\nu}}$ and the ADM decomposition
\beqa
^{4}g_{\mu\nu}&=&\left[\begin{array}{cc}-(N^{2}-N_{a}N^{a}) & N_{a} \\ N_{b} & ^{3}g_{ab} \end{array}\right]\\
^{4}g^{\alpha\beta}&=&\left[\begin{array}{cc}-\frac{1}{N^{2}} & \frac{N^{a}}{N^{2}}\\
\frac{N^{b}}{N^{2}}& ^{3}g^{ab}-\frac{N^{a}N^{b}}{N^{2}}\end{array}\right]
\eeqa
we can relate the three dimensional fields to the three-metric and the lapse and shift as follows:
\beqa
\eta_{IJ}e^{I}_{a}e^{J}_{b}&=& ^{3}g_{ab} \nn\\
t_{I}e^{I}_{a} &=& N_{a}\nn\\
t_{I}t^{I}&=&-(N^{2}-N_{a}N^{a})
\eeqa
and
\beqa
\eta^{IJ}\theta^{a}_{I}\theta^{b}_{J}&=&^{3}g^{ab}-\frac{N^{a}N^{b}}{N^{2}} \nn\\
\theta^{I}\theta^{b}_{I}&=&\frac{N^{b}}{N^{2}}\nn\\
\theta_{I}\theta^{I}&=&-\frac{1}{N^{2}} \ .
\eeqa
\section{Characterizing the degeneracies of the symplectic form\label{AppendixB}}
The unreduced symplectic form, $\bOmega=\int_\Sigma \PS\dl \omega \w \,\dl (e\,e)$, that emerges from the action principle is strictly speaking only pre-symplectic---that is, whereas it is an anti-symmetric two form ($\bm{\Omega(\bar{X},\bar{Y})}=-\bm{\Omega(\bar{Y},\bar{X})}$), and it is closed ($\dl\bOmega=0$), it is {\it degenerate}. In other words, there exist non-zero vector fields $\bm{\bar{Z}}\in T\Gamma_H$ such that $\bm{\Omega(\bar{Z},\ )}=0$. We would like to characterize this degeneracy in more detail. To this end, we write the vector field, $\bm{\bar{Z}}$, in component notation: 
\beq
\bm{\bar{Z}}=\int_{\Sigma}Z^{(e)}\frac{\dl}{\dl e}+Z^{(\omega)}\frac{\dl}{\dl \omega}
\eeq
where $Z^{(e)}=\frac{1}{2}\gamma_{I}\,(Z^{(e)})^{I}_{a}\,dx^{a}$ and
$Z^{(\omega)}=\frac{1}{4}\gamma_{I}\gamma_{J}\,(Z^{(\omega)})^{[IJ]}_{a}\,dx^{a}$. The condition $\bm{\Omega}(\bm{\bar{Z}},\
)=0$ can then be written
\beqa
e\,Z^{(e)}+Z^{(e)}\,e &=& 0\nn\\
P_{\star}Z^{(\omega)}\,e+e\,P_{\star}Z^{(\omega)}&=& 0\,.
\eeqa
Using the inverse properties of $e$ described in appendix \ref{Inverses}, we can solve the first of these equations to give $Z^{(e)}=0$. However, simply counting the degrees of freedom reveals that the second equation cannot restrict $Z^{(\omega)}$ completely. This equation has $12$ independent degrees of freedom whereas $Z^{(\omega)}$ has $18$. The maximum amount of information we can glean from this constraint is\footnote{Although these two equations appear to yield $3\times 4+4=16$ constraints on $(Z^{(\omega)})^{[IJ]}_a$, several of the constraints are not independent of each other, and taking this into account they do, in fact, produce the appropriate $12$ constraints on $(Z^{(\omega)})^{[IJ]}_a$.}
\beqa
\theta_{I}{P^{IJ}_{\star}}_{KL}\,(Z^{(\omega)})^{KL}_{a}&=&0 \nn\\
\theta^a_I{P^{IJ}_{\star}}_{KL}\,(Z^{(\omega)})^{KL}_{a}\,&=&0\,,
\eeqa
so we still have $18-12=6$ degrees of freedom left. Thus, we conclude that $\bm{\bar{Z}}$ generates a (pre)symplecto-morphism ($\bm{\mathcal{L}_{\bar{Z}}\Omega}=0$) with $6$ local degrees of freedom. Furthermore, the degeneracy only generates transformations in the ``$\omega$" direction, a reflection of the fact that it is $\omega$ that has spurious degrees of freedom that need to taken into consideration or factored away. 

Far from being a simple nuisance, the degenerate directions of $\bOmega$ are simply a reflection of the peculiar form of the Einstein-Cartan equations of motion, and are therefore an essential component to the Hamiltonian formulation therein. Consider, for example, the Einstein Cartan equations of motion (\ref{EOM1}) and (\ref{EOM2}), repeated here:
\beqa
P_{\star}R\,\vep-\vep\,P_{\star}R-{\ts\frac{2\Lambda}{3}}\star \vep\,\vep\,\vep&=&0 \\
D(P_{\star}\vep\,\vep)&=& 0 \,.
\eeqa
It can be shown \cite{Randono:CanonicalLagrangian} that Hamilton's equation of motion, $\bm{\Omega(\bar{t},\ )}=\dl C_{tot}(t,\lambda)$, is precisely the time time components of these equations pulled back to the spatial hypersurface:
\beqa
\sigma^* \left(i_{\bar{t}}\left(P_{\star}R\,\vep-\vep\,P_{\star}R-{\ts\frac{2\Lambda}{3}}\star \vep\,\vep\,\vep\right)\right)&=&0 \\
\sigma^*\left(i_{\bar{t}}\left(D(P_{\star}\vep\,\vep)\right)\right)&=& 0 \,.
\eeqa
Writing this out explicitly, we have
\beqa
\PS \mathcal{L}_{\bar{t}}\omega\,e+e\,\PS \mathcal{L}_{\bar{t}}\omega&=& -\PS D\lambda \,e+e\,\PS D\lambda +[t,\PS(R-\Lambda\, e\,e)] \nn\\
\mathcal{L}_{\bar{t}}e\,e+e\,\mathcal{L}_{\bar{t}}e &=& D[t,e]+[\lambda , e\,e]\,.
\eeqa
From these equations alone, it is clear from the discussion above that one can solve for the components $\mathcal{L}_{\bar{t}}e$ uniquely (once $t$ and $\lambda$ are specified); however, given a solution $\mathcal{L}_{\bar{t}}\omega$ to the above equations, $\mathcal{L}_{\bar{t}}\omega+Z^{(\omega)}$ also is a solution. Thus, the components $\mathcal{L}_{\bar{t}}\omega$ are only unique up to addition of a vector in the kernel of $\bOmega$. Such is the nature of the Einstein-Cartan equations of motion when restricted to a spatial slice, and the degeneracy of the symplectic structure is simply a reflection of this fact.
\end{appendix}
\bibliography{GaugeFreeV4Bib}

\end{document}